\documentclass[%
 reprint,
 amsmath,amssymb,
]{revtex4-2}

\usepackage{graphicx}
\usepackage{dcolumn}
\usepackage{bm}
\usepackage{ulem}

\usepackage{amssymb} 
\usepackage{multirow}
\usepackage{booktabs}
\usepackage{color}     
\usepackage{amsmath}
\usepackage{multirow}  

\usepackage{hyperref} 
\hypersetup{
            colorlinks=true,
            linkcolor=blue,
            filecolor=magenta,      
            urlcolor=cyan, 
            citecolor=red,}
            
\begin{document}

\preprint{APS/123-QED}

\title{Systematic study of the half-lives of nuclear bound-state $\beta^-$ decay} 

\author{Jing-Wen Ran}%
\affiliation{School of Physical Science and Technology, Southwest University, Chongqing 400715, China}%

\author{Long-Jun Wang}
\email{longjun@swu.edu.cn}
\affiliation{School of Physical Science and Technology, Southwest University, Chongqing 400715, China} 

\date{\today}

\begin{abstract}
Nuclear bound-state $\beta^-$ decay ($\beta_{\text b}$ decay) is a novel weak-interaction process that becomes possible when atoms are highly ionized, such as in stellar environments or heavy‑ion storage rings. In this work we present a systematic theoretical calculations for the $\beta_{\text b}$-decay half-lives of interesting candidates for the first time, where both allowed Gamow-Teller transitions and first-forbidden transitions are taken into account by the microscopic projected shell model, and the lepton phase space is calculated by the Takahashi-Yokoi model. We analyzed the structure informations for hundreds of nuclei near the $\beta$-stability line, and select 16 interesting candidates belonging to two categories, i.e., nuclei with negative $Q$ values and positive $Q$ values in neutral atoms respectively. Among these candidates, we recommend $^{243}\mathrm{Am}^{95+}$, $^{194}\mathrm{Os}^{76+}$, $^{227}\mathrm{Ac}^{89+}$, $^{228}\mathrm{Ra}^{88+}$, $^{241}\mathrm{Pu}^{94+}$, $^{247}\mathrm{Cm}^{96+}$ and $^{250}\mathrm{Cm}^{96+}$ as promising ones for future studies of storage-ring experiments because their $\beta_{\text{b}}$-decay half-lives are predicted to be much shorter than the half-lives in neutral atoms. These findings provide essential nuclear inputs for astrophysical models and identify specific candidates where experimental verification would be most valuable. 
\end{abstract}

\maketitle


\section{\label{sec:intro}Introduction}

Nuclear bound-state $\beta^-$ decay is one of the novel weak-interaction processes. Different from the continuum $\beta^-$ decay ($\beta_{\text{c}}$ decay, as shown in Eq. (\ref{eq.schematic}a)) where the electron and antineutrino are emitted into the continuum states, in the bound-state $\beta^-$ decay ($\beta_{\text{b}}$ decay, see Eq. (\ref{eq.schematic}b)) provided that atoms get highly ionized to be hydrogen-like or even bare nuclei, the electron is not emitted to the continuum but occupying directly the bound states of electron orbitals of daughter atoms \cite{Litvinov_2011_Rep_Prog_Phys}. 
\begin{subequations} \label{eq.schematic}
\begin{eqnarray} 
  ^{A}_{Z}\text{X}_{N}^{0+}  &\rightarrow& ^{\ \ \ A}_{Z+1}\text{Y}_{N-1}^{1+} + e^- + \tilde{\nu}_e , \\
  ^{A}_{Z}\text{X}_{N}^{Z+}  &\rightarrow& ^{\ \ \ A}_{Z+1}\text{Y}_{N-1}^{Z+}       + \tilde{\nu}_e , 
\end{eqnarray}
\end{subequations}

The exotic $\beta_{\text{b}}$-decay channel is possible in stellar environments with high temperature or in heavy-ion storage rings where heavy atoms get highly ionized. The nuclear $\beta_{\text{b}}$ decay is important for understanding the nature of neutrinos and the slow ($s$-) neutron-capture processes \cite{s_process_RMP_2011}. The concept of the $\beta_b$ decay was first proposed by Daudel \textit{et al.} in 1947~\cite{Daudel_first_1947}. A practical theoretical framework for calculating the $\beta_b$-decay half-life of highly ionized atoms was developed by Takahashi and Yokoi in the 1980s \cite{Takahashi_Yokoi_1983_NPA}. Over the past decades, experiments at heavy-ion storage rings were conducted and $\beta_{\text{b}}$ decays have been observed sequentially in $^{163}$Dy$^{66+}$ \cite{Jung_163Dy_PRL_1992}, $^{187}$Re$^{75+}$ \cite{Bosch_187Re_PRL_1996}, $^{207}$Tl$^{81+}$ \cite{Ohtsubo_207Tl_PRL_2005} and $^{205}$Tl$^{81+}$ \cite{Leckenby_Nature_2024, Sidhu_PRL_2024}. More important candidates of the nuclear $\beta_{\text{b}}$ decay are expected to be detected and studied, for which theoretical calculations and analysis should be meaningful. 

Theoretically, calculations of the $\beta_{\text{b}}$-decay half-lives are basically based on the Takahashi-Yokoi model \cite{Takahashi_Yokoi_1983_NPA} where nuclear transition strengths, including both allowed and forbidden transitions, are indispensable key inputs. However, in most of the relevant calculations, unknown nuclear transition strengths are not from sophisticated nuclear many-body calculations, but estimated empirically either from the inverse electron-capture (EC) data, or from the systematics of transitions of neighboring nuclei with available data (from ground states or isomers) \cite{TY_table_1987, Gupta_PRC_2019_bound_state_beta, Liu_Shuo_PRC_2021, C_Qi_EPJA_2026}. The former is limited to transitions from ground states of EC parent nuclei, while the latter prescription has been challenged by modern charge-exchange reactions \cite{BGao_2021_PRL_59Fe} or shell-model calculations \cite{Kuo_Ang_Li_2021_ApJL}. This is because most of the $\beta_{\text{b}}$-decay candidates are heavy nuclei and it is challenging to describe the allowed and forbidden transitions of heavy nuclei. 

Recently, we developed the traditional projected shell model (PSM) \cite{PSM_review, Sun_1996_Phys_Rep} to consider extended quasiparticle (qp) configuration space \cite{LJWang_2014_PRC_Rapid, LJWang_2016_PRC}, and then to describe allowed Gamow-Teller (GT) transition \cite{LJWang_2018_PRC_GT, ZRChen_PLB2024} and first-forbidden transition \cite{BLWang_1stF_2024} of nuclear $\beta$ decay, with the help of the Pfaffian and other algorithms \cite{Mizusaki_2013_PLB, ZRChen_2022_PRC, BLWang_2022_PRC}. The extended PSM method is then combined with the Takahashi-Yokoi model to provide a microscopic method to calculate the $\beta_{\text{b}}$-decay half-lives for the first time \cite{YXiao_bound_state_PRC_2024}. Accordingly, the $\beta_{\text{b}}$-decay half-lives of $^{163}$Dy$^{66+}$ and $^{187}$Re$^{75+}$ are described successfully and the half-life of $^{205}$Tl$^{81+}$ is predicted \cite{YXiao_bound_state_PRC_2024}. Further systematic calculations and research should be meaningful and beneficial for future experimental studies. In this work, we apply the PSM and Takahashi-Yokoi model to provide a systematic study of the half-lives of nuclear $\beta_{\text{b}}$ decay. By analyzing the $Q$ values and levels for hundreds of nuclei near the $\beta$-stability line on the chart of nuclides, we selected, in addition to the three examples of $^{163}$Dy$^{66+}$, $^{187}$Re$^{75+}$ and $^{205}$Tl$^{81+}$ examined in the previous study \cite{YXiao_bound_state_PRC_2024}, sixteen examples for further investigation. These examples are $^{193}\mathrm{Ir}^{77+}$, $^{194}\mathrm{Au}^{79+}$, $^{202}\mathrm{Tl}^{81+}$, $^{213}\mathrm{Po}^{84+}$, $^{215}\mathrm{At}^{85+}$, $^{222}\mathrm{Rn}^{86+}$, $^{243}\mathrm{Am}^{95+}$, and $^{246}\mathrm{Bk}^{97+}$, as well as $^{194}\mathrm{Os}^{76+}$, $^{212}\mathrm{At}^{85+}$, $^{227}\mathrm{Ac}^{89+}$, $^{228}\mathrm{Ra}^{88+}$, $^{241}\mathrm{Pu}^{94+}$, $^{247}\mathrm{Cm}^{96+}$, $^{249}\mathrm{Bk}^{97+}$, and $^{250}\mathrm{Cm}^{96+}$. Among these, $^{243}\mathrm{Am}^{95+}$, $^{194}\mathrm{Os}^{76+}$, $^{227}\mathrm{Ac}^{89+}$, $^{228}\mathrm{Ra}^{88+}$, $^{241}\mathrm{Pu}^{94+}$, $^{247}\mathrm{Cm}^{96+}$ and $^{250}\mathrm{Cm}^{96+}$ are recommended for future storage-ring experiments due to their much shorter $\beta_{\text{b}}$-decay half-lives compared to the half-lives in neutral atoms. 

The paper is organized as follows. In Sec. \ref{sec:theory} we introduce briefly the basic framework for calculating the $\beta_{\text{b}}$-decay half-life of highly ionized atoms. The systematic calculations and predictions are discussed in details in Sec. \ref{sec:result}, and we finally summarize our work in Sec. \ref{sec:sum}.

\section{\label{sec:theory}Theoretical framework}

The rate of $\beta_{\text{b}}$ decay from highly ionized parent $^{A}_{Z}\text{X}_{N}^{Z+}$ to daughter $^{\ \ \ A}_{Z+1}\text{Y}_{N-1}^{Z+}$ as shown in Eq. (\ref{eq.schematic}b) can be calculated by the Takahashi-Yokoi model as \cite{Takahashi_Yokoi_1983_NPA},
\begin{eqnarray} \label{eq.lambda_b}
  \lambda^{\beta_{\text{b}}}_{\text{I}} = \sum_F \lambda^{\beta_{\text{b}}}_{\text{IF}} 
  = \sum_F \frac{\ln 2}{(ft)_{\text{IF}} } f^\ast_{\text{IF} (m)} \ , 
\end{eqnarray}
with $I$ labeling the initial state (usually ground state or isomer) and the summation over all final ($F$) states within the $Q$-value window. The $\beta_{\text{b}}$-decay half-life can be obtained by $T_{1/2}(\beta_{\text{b}}) = \ln 2 / \lambda^{\beta_{\text{b}}}_{\text{I}}$. One of the important properties of $\beta_{\text{b}}$ decay is that its $Q$ value is different from that of $\beta_{\text{c}}$ decay. For fully ionized atom (bare nucleus), one can get \cite{Litvinov_2011_Rep_Prog_Phys},
\begin{eqnarray} \label{eq.Q_value}
  Q_{\beta_{\text{b}}} (K, L, \cdots) = Q_{\beta_{\text{c}}} - \Delta B_{e^-} + B_{e^-}^{K, L, \cdots} , 
\end{eqnarray}
where $Q_{\beta_{\text{c}}}$ is the $\beta_{\text{c}}$-decay $Q$ value of neutral atoms defined as the atomic mass difference of the neutral parent and daughter atoms, and $Q_{\beta_{\text{b}}}$ is the corresponding $\beta_{\text{b}}$-decay $Q$ value of highly ionized atom. $B_{e^-}^{K, L, \cdots}$ is the binding energy of the created electron in the daughter atom depending on which ($K$-, $L$-, or other) shell is occupied. $\Delta B_{e^-} = B_n (Z+1) - B_n (Z)$ is the difference of the sum of all electron binding energies for atoms. 

In Eq. (\ref{eq.lambda_b}), the indispensable key input is the comparative half-life, $(ft)_{\text{IF}}$, in the terrestrial condition. The partial half-life, $t$, can be calculated by \cite{BLWang_1stF_2024, Zhi_FF_PRC_2013},
\begin{eqnarray} \label{eq.t_here}
  \frac{K_0}{t} 
  = \int_{1}^{Q_{\text{IF}}} C(W) F_0(Z+1, W) pW (Q_{\text{IF}} - W)^{2} dW, \nonumber \\
\end{eqnarray}
where $K_0 = 6144 \pm 2 \ s$ \cite{Hardy_2009_PRC}, $W$ and $p = \sqrt{W^2 - 1}$ label the dimensionless total energy and momentum of the electron in units of $m_e c^2$ and $m_e c$ respectively. $F_0$ is the Fermi function, and $Q_{\text{IF}}$ denotes the actual (dimensionless) $Q$ value of the specific transition as, 
\begin{eqnarray} \label{eq.Qif}
  Q_{\text{IF}} = \frac{1}{m_e c^2 }(M_p - M_d + E_I -E_F ), 
\end{eqnarray}
with $E_I (E_F)$ being the nuclear excitation energy of initial (final) state for parent (daughter) nucleus with nuclear mass $M_p (M_d)$. In Eq. (\ref{eq.t_here}) $C(W)$ is the shape factor for the transition, and the phase-space integral $f$ in Eq. (\ref{eq.lambda_b}) corresponds to the integral in Eq. (\ref{eq.t_here}) without the $C(W)$. 

The calculation of the shape factor $C(W)$ is challenging theoretically, especially for forbidden transitions of heavy deformed nuclei. In principle, $C(W)$ is determined by the matrix elements of specific operators between initial and final nuclear states, i.e., $\big\langle \Psi _{J_F, \pi_F}^{n_{F}} \big\| \hat{\mathcal O} \big\| \Psi_{J_I, \pi_I}^{n_I} \big\rangle$. For allowed GT transitions with the selection rule of the spin-parity difference as $\Delta J^{\Delta \pi} \equiv |J_I-J_F|^{\pi_I \pi_F} = 0^+, 1^+$, one kind of matrix elements contribute where the corresponding operator $\hat{\mathcal O}$ is the well-known GT operator. For first-forbidden transitions with the selection rule as $\Delta J^{\Delta \pi} = 0^-, 1^-$ and $\Delta J^{\Delta \pi} = 2^-$ for nonunique and unique first-forbidden transitions respectively, nine kinds of matrix elements contribute \cite{Behrens_NPA_1971, Suzuki_first_forbidden_PRC_2012, Zhi_FF_PRC_2013, BLWang_1stF_2024}. In this work, these matrix elements are calculated by the PSM where nuclear wave function is written in the laboratory frame with good angular momentum and parity in terms of the projected basis, 
\begin{eqnarray} \label{eq.wave_function}
  | \Psi^{n}_{JM} \rangle = \sum_{K\kappa} f_{K\kappa}^{Jn} \hat{P}_{MK}^{J} | \Phi_{\kappa} \rangle ,
\end{eqnarray}
where $\Phi_{\kappa}$ labels different orders of qp configurations in the intrinsic frame \cite{LJWang_2014_PRC_Rapid, LJWang_2018_PRC_GT} and the angular-momentum-projection operator reads as,
\begin{eqnarray} \label{AMP_operator}
    \hat{P}^{J}_{MK} = \frac{2J + 1}{8\pi^2} \int d\Omega D^{J\ast}_{MK} (\Omega) \hat{R} (\Omega) ,
\end{eqnarray}
where $D^J_{MK}$ ($\hat R$) is the Wigner $D$ function (rotation operator) with respect to the Euler angle $\Omega$ \cite{ZRChen_2022_PRC, BLWang_2022_PRC} with $M$ ($K$) being the spin projection in the laboratory (intrinsic) frame. $f$ in Eq. (\ref{eq.wave_function}) labels the expansion coefficients that can be obtained by solving the corresponding eigen equation. The projection operator transforms the description of nuclei from the intrinsic to the laboratory frame. In this work we adopt three major shell in the model space, and up to 4-qp (5-qp) configurations in the configuration space for even-mass (odd-mass) heavy nuclei. For allowed GT transitions, a 0.75 quenching factor is adopted as in the literature \cite{martinez1996, YXiao_bound_state_PRC_2024}, and no quenching factors are adopted for forbidden transitions. All the PSM model parameters are adopted as in previous studies \cite{LJWang_2018_PRC_GT, LJWang_2021_PRL, ZRChen_PLB2024, BLWang_1stF_2024, YXiao_bound_state_PRC_2024, QYHu_2025_PRL}, see \cite{BLWang_1stF_2024, YXiao_bound_state_PRC_2024} for more details.

In Eq. (\ref{eq.lambda_b}), $f^\ast$ is the lepton phase volume function in the Takahashi-Yokoi model \cite{Takahashi_Yokoi_1983_NPA},
\begin{eqnarray} \label{eq.f_star}
  f^\ast_{\text{IF} (m)} = \sum_x \sigma_x (\pi / 2) \left[ g_x \text{ or } f_x \right]^2 q^2 \mathcal S_{(m)x},
\end{eqnarray}
with $m=$ a, nu, u labeling allowed, non-unique first-forbidden and unique first-forbidden transitions, respectively. $\sigma_x$ describes the vacancy of the electron orbital $x$ which lies between zero and unity (unity is adopted here). $\left[ g_x \text{ or } f_x \right]$ is understood as the larger component of the (dimensionless) electron radial wave functions for orbital $x$ evaluated at the nuclear radius $R$, i.e., 
\begin{eqnarray}
  f_x = \frac{P(R)}{R} \lambdabar^{3/2}, \qquad g_x = \frac{Q(R)}{R}\lambdabar^{3/2} ,
\end{eqnarray}
where $\lambdabar = \hbar/m_e c$ is the reduced Compton wavelength of electron, $P$ and $Q$ are the upper- and lower-component radial functions of the electron Dirac wave function in the Coulomb field of the daughter atom, which are calculated numerically by the recent RADIAL subroutine \cite{Radial_code_CPC_2019}. 

In Eq. (\ref{eq.f_star}) $q= Q_{\beta_{\text{b}}} / m_e c^2$ is the dimensionless $Q$ value, and the spectral shape factors $\mathcal S_{(m)x}$ to the lowest order are given by \cite{Takahashi_Yokoi_1983_NPA}, 
\begin{eqnarray} \label{Smx}  
  \mathcal S_{(m)x} = 
  \left\{ \begin{array}{ll} 
      1             & \text{for } m= \text{a, nu and } x=ns_{1/2}, np_{1/2}, \\ 
      q^2           & \text{for } m= \text{u and } x=ns_{1/2}, np_{1/2}, \\ 
      \frac{9}{R^2} & \text{for } m= \text{u and } x=np_{3/2}, nd_{3/2}, \\ 
      0             & \text{otherwise}.  \\ 
  \end{array} \right. 
\end{eqnarray}

\section{\label{sec:result} Calculations and Analysis}

In this section we provide our systematic calculations of nuclear $\beta_{\text{b}}$-decay half-lives of interesting candidates. As can be seen from Eq. (\ref{eq.Q_value}) that $Q_{\beta_{\text b}}$ usually differs from $Q_{\beta_{\text c}}$. When atoms get highly ionized to be hydrogen-like or even bare nuclei in stellar environments or heavy-ion storage ring, the produced electron during $\beta_{\text b }$ decay can occupy the $K$-shell orbitals of the daughter atom. In such a case, as $B_{e^-}^{K}$ is usually larger than $\Delta B_{e^-}$, $Q_{\beta_{\text b}}$ is then usually larger than $Q_{\beta_{\text c}}$. This indicates that, on one hand, a stable atom with negative $Q_{\beta_{\text{c}}}$ would become an unstable nucleus if $Q_{\beta_{\text{b}}}$ is positive and then the $\beta_{\text{b}}$ decay channel is available in highly ionized environments. Typical examples of this type and the corresponding $Q$-value information can be seen from Table \ref{tab1}. On the other hand, for some nuclei that can $\beta_{\text{c}}$ decay with small $Q_{\beta_{\text c}}$ value, the increased $Q_{\beta_{\text{b}}}$ would enable transitions to more low-lying states of the daughter nucleus, which may lead to totally different $\beta_{\text{b}}$-decay rate compared with $\beta_{\text{c}}$-decay rate. Typical examples of this type and the corresponding $Q$-value information can be seen from Table \ref{tab2}.

Before systematic calculations, we analyzed the $Q$-value information and excitation levels for hundreds of nuclei near the $\beta$-stability line on the nuclear chart, and selected promising candidate nuclei according to the above two categories. For the first category with negative $Q_{\beta_{\text{c}}}$ but positive $Q_{\beta_{\text{b}}}$, except for the two examples of $^{163}$Dy and $^{205}$Tl that have been studied in Ref. \cite{YXiao_bound_state_PRC_2024}, as well as the two special examples of $^{148}$Eu and $^{244}$Pu for which the corresponding transitions are fifth forbidden transitions so that the $\beta_{\text{b}}$ decay should be negligible, all the remaining candidates are listed in Table \ref{tab1}. The second category consists of nuclei that are already $\beta^-$-unstable in the neutral atoms ($Q_{\beta_{\text c}}>0$), for which the $\beta_{\text b}$ channel provides additional decay paths to low-lying states of the daughter nucleus. In this case, since for large $Q_{\beta_{\text c}}$ value, the $\beta_{\text b}$ decay channel would be negligible compared with the $\beta_{\text c}$ decay channel because the latter has a much larger lepton phase space than the former, accordingly, we restrict our selection in this category to nuclei with $Q_{\beta_{\text c}} \lesssim 200$ keV and ensuring that more transitions to low-lying states become possible when the $Q$‑value increases from $Q_{\beta_{\text c}}$ to $Q_{\beta_{\text b}}$. Except for the typical example of $^{187}$Re that has been studied in Ref. \cite{YXiao_bound_state_PRC_2024}, the corresponding heavy nuclei are then listed in Table \ref{tab2}. As can be seen from Tables \ref{tab1} and \ref{tab2}, most of the candidate nuclei are deformed heavy nuclei \cite{moller2016} for which the theoretical description of corresponding allow GT and first-forbidden transitions are challenging.


\begin{table}
  \caption{$Q$ values (in keV) for continuum ($\beta_{\text c}$) and bound-state ($\beta_{\text b}$) $\beta^-$ decays of nuclei with negative $Q_{\beta_{\text c}}$ values, which are stable against $\beta_{\text c}$ decay but become unstable when the $\beta_{\text{b}}$ decay channel is available in highly ionized environments. The corresponding atomic masses and electron binding energies are taken from Refs. \cite{AME2020_CPC_2021, Atom_binding_data} respectively. See the text for details. }    
  \label{tab1}
\begin{ruledtabular}
\begin{tabular}{ccccc}
  Decays & $Q_{\beta_{\text{c}}}$ & $\Delta B_{e^-}$ & $B_{e^-}^{K}$ & $Q_{\beta_{\text{b}}} (K)$  \\ 
  (g.s.$\rightarrow$g.s.) &  (keV)  &  (keV)  &  (keV)  &  (keV)   \\ \hline
  $^{193}$Ir $\rightarrow$ $^{193}$Pt  &  -56.6 & 15.9 & 90.660 & 18.160  \\
  $^{194}$Au $\rightarrow$ $^{194}$Hg  &  -27.9 & 16.8 & 95.898 & 51.198   \\
  $^{202}$Tl $\rightarrow$ $^{202}$Pb  &  -39.4 & 17.3 & 101.337& 44.637   \\
  $^{213}$Po $\rightarrow$ $^{213}$At  &  -74.0 & 18.4 & 109.887 & 17.487  \\
  $^{215}$At $\rightarrow$ $^{215}$Rn  &  -88.0 & 18.7 & 112.842 & 6.142  \\
  $^{222}$Rn $\rightarrow$ $^{222}$Fr  &  -6.0 & 19.2 & 115.858 & 90.658   \\
  $^{243}$Am $\rightarrow$ $^{243}$Cm  &  -6.9 & 24.0 & 145.740 & 114.840   \\
  $^{246}$Bk $\rightarrow$ $^{246}$Cf  & -120.2& 23.0 & 153.124 & 9.924  \\
\end{tabular}
\end{ruledtabular}
\end{table}

\begin{table}
  \caption{The same as Table \ref{tab1} but for the interesting nuclei with positive $Q_{\beta_c}$ values. See the text for details. }    
  \label{tab2}
\begin{ruledtabular}
\begin{tabular}{ccccc}
  Decays & $Q_{\beta_{\text{c}}}$ & $\Delta B_{e^-}$ & $B_{e^-}^{K}$ & $Q_{\beta_{\text{b}}} (K)$  \\ 
  (g.s.$\rightarrow$g.s.) &  (keV)  &  (keV)  &  (keV)  &  (keV)   \\ \hline
  $^{194}$Os $\rightarrow$ $^{194}$Ir  &  96.6 & 15.6 &   88.114 & 169.114  \\
  $^{212}$At $\rightarrow$ $^{212}$Rn  &  30.9 & 18.7 &  112.842 & 125.042   \\
  $^{227}$Ac $\rightarrow$ $^{227}$Th  &  44.7 & 20.0 &  125.250 & 149.950   \\
  $^{228}$Ra $\rightarrow$ $^{228}$Ac  &  45.5 & 20.0 &  122.063 & 147.563   \\
  $^{241}$Pu $\rightarrow$ $^{241}$Am  &  20.8 & 22.0 & 142.154  & 140.953   \\
  $^{247}$Cm $\rightarrow$ $^{247}$Bk  &   43.0& 22.0 &  149.398 & 170.398   \\
  $^{249}$Bk $\rightarrow$ $^{249}$Cf  &  123.6& 23.0 & 153.124  & 253.724  \\
  $^{250}$Cm $\rightarrow$ $^{250}$Bk  &   38.0& 22.0 & 149.398  & 165.398  \\ 
\end{tabular}
\end{ruledtabular}
\end{table}

\begin{table*}
  \caption{The decay information of the candidates in the first category as shown in Table \ref{tab1}, i.e., $^{193}$Ir, $^{194}$Au, $^{202}$Tl, $^{213}$Po, $^{215}$At, $^{222}$Rn, $^{243}$Am, and $^{246}$Bk, in their neutral atoms and bare nuclei, respectively. The second and third columns show the decay mode and the corresponding half-life in neutral atoms, and the fifth and sixth columns show the corresponding transitions which are available in the $\beta^-_{\text b}$ decay channel, where all the experimental data are taken from the evaluated NNDC databases \cite{NNDC}. The last two columns show the calculated log$ft$ values and the corresponding $\beta_{\text b}$ decay half-lives by the PSM and Takahashi-Yokoi model. See the text for details. }
  \label{tab3}
\begin{ruledtabular}
\begin{tabular}{cccccccccc}
 Nuclei& Decay mode& $T_{1/2}$ &$ \beta^- $ decay& \multicolumn{3}{c}{$\beta^-_{\text b}$ transition} & Type & \multicolumn{2}{c}{Theo. (PSM)}\\ \cline{5-7}\cline{9-10}
    &(neutral) &(neutral) & daughter nuclei  & [E(keV), $J^\pi_I$] & $\rightarrow$& [E(keV), $J^\pi_F$]  & (bare)& Log$ft$ & $T_{1/2}(\beta^-_{\text{b}})$\\ 
  \hline

\multirow{3}{*}{$^{193}$Ir  }& \multirow{3}{*}{stable} & \multirow{3}{*}{--- } &\multirow{3}{*}{$^{193}$Pt }
    &\multirow{3}{*}{[0.0, $3/2^+$]} & \multirow{3}{*}{$ \rightarrow$} & [0.0, $ 1/2^-$ ]&  \multirow{3}{*}{nu} & 9.155 & \multirow{3}{*}{198.30 yr}  \\
   & & & & & & [1.6, $ 3/2^-$ ] &   & 7.727 &  \\
   & & & & & & [14.3, $ 5/2^-$ ] &   & 6.849 &  \\ 
  \hline   

\multirow{1}{*}{$^{194}$Au} & $\varepsilon $, $ \beta^+ $ & 38.06 h&$^{194}$Hg 
    & [0.0, $1^-$] &$\rightarrow$&[0.0, $0^+$] & nu &6.429  &  \multirow{1}{*}{1.30 yr}   \\ 
  \hline   

\multirow{1}{*}{$^{202}$Tl}& $\varepsilon $ & 12.4706 d&$^{202}$Pb
    & [0.0, $2^-$]& $\rightarrow$& [0.0, $ 0^+$] &  u& 10.787 & $4.40 \times 10^{6}$ yr  \\

  \hline   

    $^{213}$Po &$\alpha $ & 3.706 $\mu $s &$^{213}$At
    & [0.0, $9/2^+$]& $\rightarrow$ & [0.0, $ 9/2^-$] & nu   & 6.153 & 4.04 yr  \\
  \hline  

   $^{215}$At  &$\alpha $ & 37 $\mu $s&$^{215}$Rn 
    &[0.0, $9/2^-$]& $\rightarrow$ & [0.0, $ 9/2^+$] & nu & 5.949 & 18.97 yr  \\ 
  \hline  

   $^{222}$Rn  &$\alpha $ & 3.82146 d &$^{222}$Fr
    & [0.0, $0^+$] &$\rightarrow$&[0.0, $2^-$] & u  & 11.220 &  $4.79 \times 10^{5}$ yr   \\ 
  \hline  

   \multirow{6}{*}{$^{243}$Am }&\multirow{6}{*}{ $\alpha $, $SF$ }&\multirow{6}{*}{7345 yr}&\multirow{6}{*}{$^{243}$Cm }
    &  \multirow{6}{*}{[0.0, $5/2^-$] } & \multirow{6}{*}{$\rightarrow$} &[0.0, $ 5/2^+$] & nu  & 6.777 & \multirow{6}{*}{55.13 d}  \\
    & & & & & & [42.0, $7/2^+$] & nu  &  7.593 &  \\
    & & & & & & [87.4, $1/2^+$] & u  &  13.544 &  \\ 
    & & & & & & [94.0, $9/2^+$] & u  &  9.196 &  \\
    & & & & & & [94.0, $3/2^+$] & nu  &  6.640 &  \\
    & & & & & & [114.0, $7/2^+$] & nu  &  6.957 &  \\   
  \hline  

    $^{246}$Bk  &$\varepsilon $, $ \beta^+ $ & 1.8 d & $^{246}$Cf 
    &[0.0, $2^-$] &$\rightarrow$&[0.0, $0^+$] & u & 10.398 & $2.11 \times 10^{8}$ yr  \\ 
\end{tabular}
\end{ruledtabular}
\end{table*}

\begin{table*}
  \caption{The same as Table \ref{tab3} but for the candidates in the second category as shown in Table \ref{tab2}, i.e., $^{194}$Os, $^{212}$At, $^{227}$Ac, $^{228}$Ra, $^{241}$Pu, $^{247}$Cm, $^{249}$Bk, and $^{250}$Cm. The last three columns show the calculated log$ft$ values and compared with available experimental data in the NNDC databases \cite{NNDC}, as well as the corresponding $\beta_{\text b}$ decay half-lives by the PSM and Takahashi-Yokoi model. See the text for details. }
  \label{tab4}

\begin{ruledtabular}
\begin{tabular}{ccccccccccc}
Nuclei & Decay mode& $T_{1/2}$ &$ \beta^- $ decay& \multicolumn{3}{c}{ $ \beta^-_{\text b}$ transition} & Type &\multicolumn{2}{c}{Log$ft$}&$T_{1/2}(\beta^-_{\text{b}}) $\\ \cline{5-7}\cline{9-10}
    &(neutral) &(neutral) & daughter nuclei  & [E(keV), $J^\pi_I$] & $\rightarrow$& [E(keV), $J^\pi_F$]  & (bare)&Exp.&PSM&(bare)\\ 
  \hline

  \multirow{8}{*}{ $^{194}$Os } & \multirow{8}{*}{$\beta^-$ }& \multirow{8}{*}{6 yr} &  \multirow{8}{*}{$^{194}$Ir}
 &\multirow{8}{*}{ [0.0, $0^+$]}& \multirow{8}{*}{ $\rightarrow$} &[0.0, $1^-$] & nu & 7.6 &6.345 & \multirow{8}{*}{65.01 d} \\ 
 &&&&& &[43.1, $0^-$] & nu & 6.3 &9.008  &  \\ 
 &&&&& &[82.3, $1^-$] & nu & 7.4 & 6.500  & \\ 
 &&&&& &[84.3, $2^-$] & u & --- & 9.913  & \\
 &&&&& &[112.2, $2^-$] & u  & --- & 10.530 & \\ 
 &&&& && [138.7, $1^-$] & nu & ---& 7.906 & \\ 
 &&&& &&[143.6, $0^-$] & nu & --- & 8.707 & \\
 &&&&& &[161.0, $1^-$] & nu & --- & 6.677 & \\ 
  \hline  

   \multirow{1}{*}{ $^{212}$At }  & $\alpha$&0.314 s & $^{212}$Rn
  & [0.0, $1^-$] &$\rightarrow$& [0.0, $0^+$] & nu & ---  & 9.716 & 267.10 yr \\
 \hline  

  \multirow{7}{*}{ $^{227}$Ac } &\multirow{7}{*}{$\beta^-$, $\alpha $}&\multirow{7}{*}{21.7725 yr} & \multirow{7}{*}{$^{227}$Th }
 & \multirow{7}{*}{[0.0, $3/2^-$]} & \multirow{7}{*}{$\rightarrow$ }&[0.0, $1/2^+$]  & nu & $\approx$7.1 & 8.232 & \multirow{7}{*}{23.45 d}  \\ 
 &&&& && [9.3, $5/2^+$]  & nu & $\approx$7.0 & 6.814&  \\ 
 &&&& && [24.4, $3/2^+$] & nu & $\approx$6.8 & 6.139 & \\ 
 &&&& && [37.9, $3/2^-$] & a  & 6.9 & 7.325& \\ 
 &&&& &&[77.6, $3/2^+$] & nu & --- & 7.820 & \\ 
 &&&& &&[99.2, ($5/2^+$)] & nu & --- & 6.332 & \\
 &&&& &&[127.3, ($3/2^+$)] & nu & --- & 6.852 & \\  
  \hline   

  \multirow{4}{*}{ $^{228}$Ra }& \multirow{4}{*}{$\beta^-$}& \multirow{4}{*}{5.75 yr} & \multirow{4}{*}{$^{228}$Ac }
 &  \multirow{4}{*}{[0.0, $0^+$] }& \multirow{4}{*}{$\rightarrow$}& [6.3, $1^-$] & nu & $\approx$7.1 &  5.958 & \multirow{4}{*}{2.02 d}  \\ 
 &&&& && [6.7, $1^+$]  & a  & $\approx$6.5 &  8.039  &  \\ 
 &&&& && [20.2, $1^-$] & nu & 6.2 &  7.353  & \\ 
 &&&& && [33.1, $1^+$] & a  & 5.12&   5.630  & \\ 
  \hline  

  \multirow{3}{*}{ $^{241}$Pu } &\multirow{3}{*}{$\beta^-$, $\alpha $}& \multirow{3}{*}{14.329 yr} & \multirow{3}{*}{ $^{241}$Am } 
 & \multirow{3}{*}{ [0.0, $5/2^+$]}& \multirow{3}{*}{ $\rightarrow$}& [0.0, $5/2^-$]  & nu & 5.788  & 6.461 & \multirow{3}{*}{4.36 d}  \\ 
 &&&& &&[41.2, $7/2^-$] & nu & ---  & 7.000  &  \\ 
 &&&& && [93.7, $9/2^-$] & u  & --- & 11.562  & \\ 
  \hline  

  \multirow{5}{*}{ $^{247}$Cm}  &\multirow{5}{*}{$\alpha $}& \multirow{5}{*}{1.56 $\times 10^{7}$ yr} & \multirow{5}{*}{$^{247}$Bk } 
 &\multirow{5}{*}{[0.0, $9/2^-$]}&  \multirow{5}{*}{$\rightarrow$ }&[40.8, $7/2^+$]  & nu & --- & 6.242  & \multirow{5}{*}{9.52 d}  \\ 
 &&&& && [71.6, $7/2^-$]  & a  & --- & 11.943 &  \\ 
 &&&& && [82.8, $9/2^+$]  & nu & --- & 6.848  & \\ 
 &&&& && [125.5, $9/2^-$] & a  & --- & 12.014 & \\ 
 &&&& && [137.0, $11/2^+$]  & nu  & --- & 7.806  & \\ 
   \hline   

  \multirow{5}{*}{ $^{249}$Bk}& \multirow{5}{*}{$\beta^-$, $\alpha$, $SF$}& \multirow{5}{*}{327.2 d} &  \multirow{5}{*}{ $^{249}$Cf }  
 &  \multirow{5}{*}{[0.0, $7/2^+$]}& \multirow{5}{*}{ $\rightarrow$}& [0.0, $9/2^-$]   & nu & 7.019 & 6.133  & \multirow{5}{*}{15.75 d}  \\ 
 &&&& &&[62.5, $11/2^-$] & u  & --- & 10.937 &  \\
 &&&& &&[145.0, $5/2^+$]   & a  & --- & 12.319 &  \\ 
 &&&& && [188.0, $7/2^+$]   & a  & --- & 6.398  & \\ 
 &&&& && [243.1, $9/2^+$] & a  & --- & 7.236  &  \\
  \hline  

  \multirow{4}{*}{ $^{250}$Cm }& \multirow{4}{*}{$SF$, $\beta^-$, $\alpha$}& \multirow{4}{*}{$\approx$8300 yr} &  \multirow{4}{*}{ $^{250}$Bk }
 & \multirow{4}{*}{[0.0, $0^+$] }& \multirow{4}{*}{$\rightarrow$ }&[0.0, $2^-$]  & u  & --- & 10.245  & \multirow{4}{*}{3.86 d}  \\ 
 &&&& && [103.8, $1^-$]  & nu & --- & 5.070  &  \\ 
 &&&& &&[125.0, $2^-$] & u  & --- & 8.666  & \\ 
 &&&& &&[146.4, $2^-$] & u  & --- & 10.203  & \\
\end{tabular}
\end{ruledtabular}
\end{table*}

We show in Table \ref{tab3} the decay information of the candidates in the first category as shown in Table \ref{tab1}. The second column of Table \ref{tab3} shows the decay mode in neutral atoms, including electron capture ($\varepsilon$), $\beta^+$ decay ($\beta^+$), $\alpha$ decay ($\alpha$) and spontaneous fission (SF), and the corresponding half-lives in neutral atoms are shown in the third column. When getting highly ionized, the $\beta_{\text b}$-decay channel becomes available and the corresponding transition information, including the excitation energies and spin-parity assignments of initial and final states, are shown in the the fifth and sixth columns. These transitions correspond to first-forbidden transitions from the ground state of the parent nucleus to the ground state and low-lying excited states of the daughter nucleus within the $Q_{\beta_{\text b}}$ window, where the excitation energies and spin-parity assignments are taken from the NNDC databases \cite{NNDC}. Accordingly, the log$ft$ values for these transitions have not been measured experimentally so that we provide the calculations within our PSM model. The $\beta_{\text b}$-decay half-lives evaluated by the Takahashi-Yokoi model are shown in the last column. 

It is seen from Table \ref{tab3} that the $\beta_{\text b}$-decay half-lives are usually different from the corresponding half-lives in neutral atoms. Experimentally, more interest lies in cases and examples where the $\beta_{\text b}$-decay half-life is much shorter than that of the neutral atom. Actually, the $\beta_{\text b}$-decay half-life is mainly determined by the $Q_{\beta_{\text b}}$ value and the $ft$ value, as can be seen from Eqs. (\ref{eq.lambda_b}, \ref{eq.f_star}). The larger the $Q_{\beta_{\text b}}$ value and/or the smaller the log$ft$ value, the shorter the half-life. As shown in Tables \ref{tab1} and \ref{tab3}, for $^{193}$Ir which is stable in neutral atom, since the $Q_{\beta_{\text b}}$ value is as small as 18.16 keV and the log$ft$ values are predicted to be $\gtrsim 7$, the $\beta_{\text b}$-decay half-life is as large as 198.3 years. The unique first-forbidden transition is usually very weak with a large log$ft$ value, as shown in the three examples $^{202}$Tl, $^{222}$Rn and $^{246}$Bk in Table \ref{tab3}, their log$ft$ values are indeed large (log$ft \gtrsim 10$), which lead to that the $\beta_{\text b}$-decay half-lives of these three nuclei are longer than $10^{5}$ years. For $^{194}$Au, $^{213}$Po and $^{215}$At, although the corresponding first-forbidden transitions are very strong with log$ft \approx 6$, and their corresponding $\beta_{\text b}$-decay half-lives associated with these transitions range from 1 to 20 years, the half-lives in neutral atoms by electron capture, $\beta^+$ decay or $\alpha$ decay are much shorter, ranging from hours down to microseconds. These findings indicate that the above seven candidates in the first category, i.e., $^{193}$Ir, $^{194}$Au, $^{202}$Tl, $^{213}$Po, $^{215}$At, $^{222}$Rn and $^{246}$Bk should not be considered promising for future experimental studies as their $\beta_{\text b}$-decay half-lives are either too long or much longer than the corresponding half-lives in neutral atoms. 

A special candidate is $^{243}$Am, which has the largest $Q_{\beta_{\text b}}$ value in the first category with $Q_{\beta_{\text b}} = 114.84$ keV as shown in Table \ref{tab1}. In addition, the corresponding ground-state to ground-state transition that has the largest lepton phase space is predicted to be strong with log$ft = 6.777$. These two aspects result in a very short $\beta_{\text b}$-decay half-life of about 55 days. By comparison, in neutral atoms, the half-life via $\alpha$ decay and spontaneous fission is as long as $7345$ years, which is longer than the predicted $\beta_{\text b}$-decay half-life by about five orders of magnitude. Therefore, we recommend $^{243}$Am as a promising candidate for future experimental studies of nuclear $\beta_{\text b}$ decay.


We now discuss the second category. Table \ref{tab4} lists the decay information of the eight candidates in the second category, i.e., $^{194}$Os, $^{212}$At, $^{227}$Ac, $^{228}$Ra, $^{241}$Pu, $^{247}$Cm, $^{249}$Bk and $^{250}$Cm, for which the $Q$-value information is shown in Table \ref{tab2}. For these candidate nuclei, $Q_{\beta_{\text b}}$ is larger than $Q_{\beta_{\text c}}$, so more transitions can contribute to $\beta_{\text b}$ decay because of the enlarged $Q$ value. In Table \ref{tab4}, we have included all transitions within the $Q_{\beta_{\text b}}$-value window, except for $^{228}$Ra. For $^{228}$Ra, its daughter nucleus is an odd-odd nucleus, and there are dozens of excited states with excitation energies lower than the corresponding $Q_{\beta_{\text b}}$ value. However, the transitions to these excited states are predicted by our PSM calculations to be all weak; their log$ft$ values are predicted to be much larger than that of the transitions to the very low‑lying states (which are about $7$ as can be seen from Table \ref{tab4}). Therefore, their contributions to the corresponding $\beta_{\text b}$-decay half‑life are negligible and they are consequently not listed in Table \ref{tab4}.

For these candidates in the second category, the $Q_{\beta_{\text c}}$ values are positive so that some transitions of the corresponding $\beta_{\text c}$ decay are possible. The experimental data of log$ft$ values of these transitions are shown in the eighth column of Table \ref{tab4}, and it is seen from Table \ref{tab4} that these data are described reasonable by our PSM calculations which are listed in the ninth column of Table \ref{tab4}. In the last column of Table \ref{tab4} we show the calculated $\beta_{\text b}$-decay half-lives where experimental log$ft$ values are adopted whenever available. As one can see from Table \ref{tab4}, the $\beta_{\text b}$-decay half-lives are shorter than the corresponding half-lives in neutral atoms for these candidates except for $^{212}$At. For $^{194}$Os, $^{227}$Ac, $^{228}$Ra, $^{241}$Pu, and $^{249}$Bk, the experimental log$ft$ values for transitions to the ground states or very low-lying states could be as large as about $6$, so that the corresponding $\beta_{\text b}$-decay half-lives are determined by these transitions with experimental log$ft$ values. The corresponding $\beta_{\text b}$-decay half-lives are on the order of days, while those of other decay modes in neutral atoms are on the order of years. Therefore, these nuclei are of great interest for future experimental studies of nuclear $\beta_{\text b}$ decay. More interestingly, $^{247}$Cm and $^{250}$Cm exhibit distinctly different behavior. As transuranium nuclei, their decay in neutral atoms is dominated by $\alpha$ decay and spontaneous fission because the corresponding $Q_{\beta_{\text c}}$ is as small as about $40$ keV as seen from Table \ref{tab2}. However, under highly ionization (i.e., to be hydrogen-like or even bare nuclei), the $\beta_{\text b}$-decay channel opens and rapidly becomes the dominant decay mode. Since the log$ft$ values of transitions to the low-lying states are predicted to be as large as about $6$ and $5$ for $^{247}$Cm and $^{250}$Cm respectively, their $\beta_{\text b}$-decay half-lives are then predicted to be on the order of days, whereas the corresponding half-lives in neutral atoms are on the order of $10^7$ and $10^4$ years respectively. This indicates a dramatic reduction in half-lives for $^{247}$Cm and $^{250}$Cm. It is noted that the half-life of $^{247}$Cm is shortened by up to nine orders of magnitude, while that of $^{250}$Cm is reduced by almost six orders of magnitude. This striking contrast reveals the profound influence of the ionization state on the decay properties of heavy nuclei: isotopes that are conventionally considered as typical $\alpha$ emitters may, in astrophysical and stellar environments, actually undergo fast $\beta_{\text b}$ decay. Given the dramatic half-life contraction predicted for these two isotopes, i.e., spanning six to nine orders of magnitude, both $^{247}$Cm and $^{250}$Cm emerge as exceptionally promising candidates for future experimental investigations of $\beta_{\text b}$ decay.

\section{\label{sec:sum}summary}

In summary, $\beta_{\text b}$ decay represents an essential decay channel for nuclei immersed in stellar plasmas or heavy-ion storage ring. In this work, we perform a comprehensive survey across the nuclear landscape to identify candidate nuclides that can undergo $\beta_{\text b}$ decay under highly ionized conditions. Interesting candidates are considered according to two distinct categories: (i) nuclides with negative $Q_{\beta_{\text c}}$ values in neutral atoms, for which the $\beta_{\text b}$ decay channel becomes energetically allowed only upon highly ionization; and (ii) nuclides are already unstable with positive $Q_{\beta_{\text c}}$ values in neutral atoms, while additional transitions are possible since the effective $Q$ value is enlarged from $Q_{\beta_{\text c}}$ to $Q_{\beta_{\text b}}$. For each category, eight representative nuclides are selected for detailed investigation. Their $\log ft$ values are calculated microscopically for the first time, by the PSM where both the allowed GT and first-forbidden transitions are treated on an equal footing. These transition strengths are then combined with phase space factors obtained from the Takahashi-Yokoi model to yield the $\beta_{\text b}$-decay half-lives.

Our calculations predict that for highly ionized $^{243}\mathrm{Am}^{95+}$, $^{194}\mathrm{Os}^{76+}$, $^{227}\mathrm{Ac}^{89+}$, $^{228}\mathrm{Ra}^{88+}$, $^{241}\mathrm{Pu}^{94+}$, $^{247}\mathrm{Cm}^{96+}$ and $^{250}\mathrm{Cm}^{96+}$, the $\beta_{\text{b}}$-decay half-lives are significantly shorter than the corresponding half-lives of other decay modes in neutral atoms. These candidates are therefore recommended for future studies of storage-ring experiments for nuclear $\beta_{\text b}$ decay. Specifically, among all the candidates studied, particular attention is drawn to $^{243}$Am, $^{247}$Cm, and $^{250}$Cm. For highly ionized $^{243}$Am, the predicted $\beta_{\text b}$ decay half-life is about 55 days, in striking contrast to its neutral-atom $\alpha$-decay half-life of 7345 years --- a reduction of nearly five orders of magnitude. For $^{247}$Cm, the half-life plunges from about $10^7$ years in the neutral state (dominated by $\alpha$ decay) to merely 0.026 years (9.52 days) under highly ionized conditions, corresponding to a compression of nine orders of magnitude. For $^{250}$Cm, the half-life decreases from about 8300 years to 0.011 years (3.86 days), a reduction of almost six orders of magnitude. These extreme cases vividly demonstrate how the ionization state can fundamentally alter the decay dynamics of heavy nuclei, transforming nuclides conventionally classified as typical $\alpha$ emitters into rapid $\beta^-$ decayers in astrophysical environments.

The present work provides the first systematic theoretical investigation of $\beta_{\text b}$ decay half-lives for a diverse set of candidate nuclei within a unified microscopic framework. The predictions provide valuable guidance for future experimental efforts, particularly those at storage-ring facilities, and establish a foundation for more realistic astrophysical simulations in which highly ionized species prevail. Future extensions of this work will incorporate the Saha equation to determine the equilibrium populations of various ionization states under stellar conditions.

\begin{acknowledgments}
  We thank Yu-Hong Zheng for checking carefully the details of the systematic calculations. This work is supported by the National Natural Science Foundation of China (Grants No. 12275225), and by the New Chongqing Youth Innovation Talent Project (Grant No. CSTB2025YITP-QCRCX0055). 
\end{acknowledgments}


%

\end{document}